\documentclass[11pt]{article}
\usepackage[T1]{fontenc}
\usepackage{lmodern}
\usepackage{graphicx}
\usepackage[margin=1in]{geometry}
\usepackage{booktabs,tabularx,array,microtype}
\usepackage{amsmath,amssymb,amsthm}
\usepackage{xurl}
\usepackage[hidelinks]{hyperref}
\usepackage[numbers,sort&compress]{natbib}
\usepackage{bm}
\usepackage{multirow}
\usepackage{xcolor,placeins,tikz}
\usepackage{float}
\usepackage{authblk}
\newfloat{algorithm}{tbp}{loa}
\floatname{algorithm}{Algorithm}
\floatstyle{ruled}
\usetikzlibrary{arrows.meta,positioning}
\restylefloat{algorithm}

\newcommand{\norm}[1]{\left \lvert\left \lvert #1 \right\rvert\right \rvert}
\newcommand{\Hi}{\mathcal{H}_\infty}
\newcommand{\RHinf}{\mathcal{R}\mathcal{H}_\infty}

\theoremstyle{definition}
\newtheorem{definition}{Definition}
\newtheorem{proposition}{Proposition}

\title{Extremal-Phase-Guided Sequential Convex Optimization:\\   Model Order Reduction}
\author[1]{Mustafa O\u{g}uz Ye\u{g}in%
\thanks{Corresponding author.
Email: \texttt{moguzyegin@gmail.com}}}
\author[2]{Hitay \"{O}zbay}
\author[1]{Tom\'{a}\v{s} Vyhl\'{i}dal}

\affil[1]{Department of Instrumentation and Control Engineering\\
Faculty of Mechanical Engineering\\
Czech Technical University in Prague\\
166 00 Prague, Czech Republic}

\affil[2]{Department of Electrical and Electronics Engineering\\
Bilkent University\\
06800 Ankara, T\"{u}rkiye}

\date{}

\begin{document}
\maketitle
\vspace{-3em}
\begin{abstract}\color{black}\emergencystretch=1em
Fixed-order frequency-weighted $\Hi$ approximation is a nonconvex problem, and local optimization methods can be sensitive to initialization. We present extremal-phase-guided sequential convex optimization (EPGSCO) for stable fixed-order rational approximation. We introduce a novel phase-guidance technique motivated by the all-pass structure underlying optimal Hankel-norm approximation, with a phase template fitted to the error-magnitude derivative at dominant peaks. We integrate this guidance with dominant-peak selection, focused frequency grids, and adaptive pole--residue and polynomial-coefficient updates to address stagnation and improve approximation accuracy. CD-player examples and benchmarks with known optimal errors evaluate the algorithm against established methods. Comparisons with classical sequential convex optimization (SCO) under shared settings for the common components assess the combined contribution of phase guidance and adaptive frequency refinement. Results demonstrate improved accuracy in most of the tested cases and attainment of the known optimum from four tested initializations for a constructed order-42 benchmark, with additional computation in some cases.
\end{abstract}


\section{Introduction}
\label{sec: introduction}
High-order rational and infinite-dimensional models arise in distributed-parameter systems, electrical circuits, power networks, and systems with communication or actuator delays \cite{Antoulas2005,Kramer2026,Aththanayake2024}. Although these models describe the underlying dynamics accurately, their direct use can be impractical for repeated simulation, optimization, or real-time implementation \cite{Benner2015}. Model-order reduction replaces them with low-order models that preserve relevant input-output properties \cite{Schilders2008,Becker2022}. The required accuracy depends on the application. Circuit models must reproduce impedance or scattering responses in relevant frequency ranges \cite{GrivetTalocia2016,Gustavsen1999}, while feedback applications depend on dynamics affecting stability, robustness, and disturbance attenuation \cite{Zhou1996}. Phase errors at specific bandwidth are particularly significant in delayed feedback systems \cite{Michiels2014}. Frequency weighting expresses this variation in importance, and the $\Hi$ criterion minimizes the largest weighted error rather than an average fitting error.

Balanced truncation removes states according to controllability and observability, preserves stability for stable systems, and provides an a priori $\Hi$ error bound \cite{Moore1981,Pernebo1982,Enns1984,Antoulas2005}. Hankel-norm approximation minimizes the Hankel norm and is associated with Glover's scaled all-pass error construction \cite{Glover1984}. Although their objectives differ from frequency-weighted $\Hi$ approximation, these methods provide well-established baselines for comparison. Moment-matching methods preserve selected interpolation conditions, while IRKA updates interpolation points to satisfy local $\mathcal H_2$-optimality conditions \cite{Grimme1997,Gugercin2008,Benner2015,Antoulas2020}. These methods are effective for large-scale problems, but the chosen interpolation data and initialization can influence the resulting approximation.

For frequency-response data, vector fitting estimates poles and residues through iterative least-squares problems \cite{Gustavsen1999}. The Adaptive Antoulas--Anderson (AAA) algorithm adaptively selects support points and constructs a barycentric rational approximant \cite{Nakatsukasa2018}. Both provide useful data-based models and initializations, although their standard formulations do not directly solve the stable fixed-order frequency-weighted minimax problem considered here. Direct optimization of that problem remains nonconvex because numerator and denominator parameters jointly determine the rational response.

Fixed-order $\Hi$ model reduction has also been studied through LMIs and semidefinite programming \cite{Helmersson1995,Sootla2012}. In a closely related approach, Shakib et al.\ combine time-domain moment matching with unweighted $\Hi$ error minimization \cite{Shakib2021}. Fixed interpolation points define a family of reduced models, whose remaining parameters are optimized through iterative LMIs. The present problem instead includes frequency weighting and updates the approximation poles without retaining a prescribed moment-matching family. Realization-based semidefinite formulations can also become computationally demanding because their constraints involve the original and reduced system dimensions. This motivates a formulation operating directly on sampled frequency responses.

Sequential convex optimization (SCO) constructs local convex subproblems and validates finite steps using the nonlinear objective \cite{Wang2006,Ren2023}. Rational-response linearizations in pole--residue or polynomial coordinates provide such subproblems. HIFOO and \texttt{systune} offer alternative fixed-order searches by treating reduced-model parameters as tunable variables and the weighted error as a performance channel \cite{Burke2006,Gumussoy2009,Apkarian2006,MathWorks2026}. HIFOO uses nonsmooth optimization based on quasi-Newton updates and gradient sampling \cite{LewisOverton2013,BurkeLewisOverton2005}. These approaches are generally sensitive to initialization and may converge to different local minima. In sampled rational approximation, local progress additionally depends on frequency coverage, coordinate conditioning, and validation: a small retained-grid error may conceal a larger error at omitted frequencies, while nearly coincident poles can impair modal updates.

In this paper, we propose extremal-phase-guided sequential convex optimization (EPGSCO) for stable fixed-order frequency-weighted $\Hi$ approximation from frequency-response data. The proposed algorithm combines dominant-error-peak selection and focused optimization grids with phase-guided joint pole--residue second-order cone programs. Second-order cone programming (SOCP) is a class of convex optimization problems with a linear objective and constraints expressed as Euclidean-norm inequalities involving affine functions \cite{Alizadeh2003}. We use this formulation to bound the magnitudes of the linearized complex error responses and to impose parameter trust regions. Scaled numerator--denominator coefficient updates are used when modal conversion fails or the pole--residue step produces insufficient improvement. Pole-frequency regularization, adaptive guidance and trust regions, and backtracking govern the reconstruction and acceptance of stable approximants. Omitted-sample checks and grid augmentation ensure that accepted steps are evaluated using the complete supplied frequency grid.

The contributions of this work are a novel extremal phase-guidance
technique and an algorithm that integrates this technique with
dominant-frequency selection, adaptive coordinate switching, and
regularized sequential convex optimization. Motivated by Glover's
all-pass error construction, we fit an all-pass template to the
magnitude-derivative phase at dominant error peaks and use the resulting directions to guide local SOCP updates. Frequency selection concentrates the optimization on relevant samples, while coordinate switching addresses numerical conditioning and insufficient progress. We adapt the phase-guidance weight before contracting the trust regions, allowing the search to depart from the fitted directions when necessary. Stability checks and nonlinear complete-grid merit evaluation govern the acceptance of trial approximants.

Weighted and unweighted CD-player comparisons evaluate the algorithm against established methods. SCO comparisons with shared settings assess the additional effects of phase guidance and adaptive frequency refinement, while constructed known-optimum benchmarks test achievable accuracy and sensitivity to initialization against a reference error level. The results demonstrate improvements in several cases, with additional computation in some runs, without establishing general global optimality. Section~\ref{sec:problem_formulation} defines the problem, Section~\ref{sec:weighted_method} develops EPGSCO, and Section~\ref{sec:numerical_examples} presents the comparisons. Appendix~\ref{app:benchmark_construction} describes the known-optimum benchmarks.

\section{Problem Formulation and Notation}
\label{sec:problem_formulation}

We consider continuous-time single-input single-output (SISO) systems with real coefficients. The notation $ j=\sqrt{-1}$ is used throughout. We assume that $G$ is stable and proper, has no essential singularity in the closed right half plane, and has a continuous frequency response on the imaginary axis. Then, $\|G\|_\infty~=~\sup_{\omega\in\mathbb R}|G(j\omega)|$. The set $\RHinf$ contains stable proper real-rational transfer functions. Frequency-response data (FRD) consist of samples on $\Omega=\{\omega_k\}_{k=1}^{N}$, ordered by increasing frequency. For real-coefficient systems, the negative-frequency samples follow by conjugation.

\subsection{Weighted rational approximation}
Let $G$ be a stable proper plant or reference transfer function, with sampled
values $G_k=G( j\omega_k)$, and let $W_k$ represent the samples of the “weighting function” $W$, where $W$ is given in the problem.
A prescribed-order ``($m$-th order)'' approximation of the $n$-th order $G$, where $n \in [(m+1),~\infty]$ is parameterized as
\begin{equation}
 Q(s;\theta)=\frac{N(s)}{D(s)},\quad
 N(s)=\sum_{\ell=0}^{m-\rho}b_\ell s^\ell,\quad
 D(s)=s^m+\sum_{\ell=0}^{m-1}a_\ell s^\ell,
 \label{eq:reduced_model_parameterization}
\end{equation}
where $D$ is Hurwitz and $\rho$ is the imposed minimum relative degree. The coefficients are real; coprimeness is needed to identify the denominator order with the McMillan degree. Cancellations or a vanishing leading numerator coefficient can reduce the realized order or increase the relative degree. The weighted approximation problem is to determine the optimal error
$\gamma_{m,\rho}$ and a corresponding optimal approximant
$Q^{\mathrm{opt}}_{m,\rho}\in\mathcal Q_{m,\rho}$, when the infimum is attained:
\begin{equation}
 \gamma_{ m,\rho}:= \inf_{Q\in\mathcal Q_{m,\rho}}\|W(G-Q)\|_\infty,
 \quad
 \mathcal Q_{m,\rho}=\{N/D:\ D\text{ Hurwitz},\ \deg D=m,
                                  \deg N\le m-\rho\}.
 \label{defn: prob_ss}
\end{equation}
An optimal approximant, when it exists, satisfies
\[
\|W(G-Q^{\mathrm{opt}}_{m,\rho})\|_\infty=\gamma_{m,\rho}~.
\]
For FRD, the implemented objective is the finite-grid surrogate
\begin{equation}
 \min_{Q\in\mathcal Q_{m,\rho}}\gamma_\Omega(Q),\qquad
 \gamma_\Omega(Q)=\max_{1\le k\le N}|E_k(Q)|,\quad
 E_k(Q)=W_k(G_k-Q_k).
 \label{defn: prob_frd}
\end{equation}
This is a nonconvex problem when numerator and denominator are variable. Its sampled optimum does not certify the continuous-frequency optimum. The weighted error must be stable and proper for the latter interpretation. The current approximation implementation supports $\rho\in\{0,1\}$.

\subsection{Iteration notation}
We use $i$ to denote the iteration index and $k$ to denote the
frequency-sample index. The set $\Omega_i$ denotes the current validation frequency grid, $\mathcal A_i$ contains the indices of selected dominant local maxima on this grid, and $\mathcal I_i$ contains the sample indices retained in the temporary SOCP grid. We call a peak active if its magnitude is within a prescribed tolerance of the current maximum; by default, this means at least $97\%$ of the maximum. $\mathcal T_i(s)$ denotes the phase polynomial. The sign $\sigma_i\in\{-1,1\}$ and cluster location $z_i>0$ parameterize the all-pass template. The nonnegative extremal weight $\lambda_i$ determines the strength of phase guidance, while $\tau_N$ and $\tau_D$ denote the trust-region radii for the numerator-coefficient or residue increments and the denominator-coefficient or pole increments, respectively. $\tau_{x,i}$ represents the trust region coefficients at $i$-th iteration, and $\tau_g$ and $\tau_s$ are the growth and shrinkage factors, respectively. The sampled objective $\gamma_\Omega$, local SOCP bound $\gamma_i+\delta\gamma$, and continuous-frequency norm $\gamma$ are distinct quantities.

\section{Extremal-Phase-Guided Weighted Approximation}
\label{sec:weighted_method}
Each iteration starts from a stable approximant $Q_i$ and follows the sequence developed below: select relevant frequency samples, form a local response model, fit phase directions, solve a guided minimax SOCP, and validate the resulting rational models. The accepted model and validated progress determine the next search settings. Algorithm~\ref{alg:weighted_epgsco}, presented at the end of this section, summarizes this procedure.

\subsection{Frequency selection}
\label{subsec:grid_switch}
We use the complete-grid residual to select up to $m+1$ dominant local maxima, including eligible endpoints. A peak-height threshold identifies dominant maxima; additional maxima are retained when needed. These selected peaks need not all attain the current maximum error.

We use $\mathcal I_i$ to denote the temporary optimization grid, which contains selected peaks, endpoints, significant-error samples, and samples found to violate a previous reduced-grid bound. Additional samples provide frequency coverage up to a target size. Significant-error samples are retained even when this exceeds that target. Grid focusing selects existing input samples; it does not evaluate new frequencies between them. The treatment of omitted-sample violations is described in Section~\ref{subsec:acceptance}.

\subsection{Local model and parameter increments}
\label{subsec:coordinates}
We use $\gamma_i=\gamma_\Omega(Q_i)$ to denote the current sampled error level. The first-order response variation $\delta Q_i$ represents the affine local model used in the SOCP. Pole--residue coordinates are used first; coefficient coordinates provide an alternative when their validated progress is insufficient.

For a model with simple poles, we use the pole--residue representation and its first variation
\begin{equation}
 Q_i(s)=d_i+\sum_{\ell=1}^{m}\frac{c_{i,\ell}}{s-p_{i,\ell}},\qquad
 \delta Q_i(s)=\delta d+\sum_{\ell=1}^{m}
 \left(\frac{\delta c_\ell}{s-p_{i,\ell}}+
 \frac{c_{i,\ell}\delta p_\ell}{(s-p_{i,\ell})^2}\right).
 \label{eq:pole_residue_derivative}
\end{equation}
We parameterize real poles and complex-conjugate pairs using real variables. Poles and residues vary jointly, with $d_i=\delta d=0$ for $\rho=1$. Nearly coincident poles can make these coordinates ill-conditioned.

For the alternative parametrization, we use numerator and denominator coefficients. For $Q_i=N_i/D_i$, differentiation gives
\begin{equation}
 \delta Q_i(s)=\frac{\delta N(s)}{D_i(s)}-
 \frac{N_i(s)\delta D(s)}{D_i(s)^2},\qquad
 \deg\delta N\le m-\rho,\quad \deg\delta D\le m-1.
 \label{eq:coefficient_derivative}
\end{equation}
The leading denominator coefficient remains fixed to preserve monicity. For coefficient scaling, we use the variable $x_i=s/\omega_{0,i}$, where $\omega_{0,i}$ is the geometric mean of the positive selected-peak frequencies. This changes the polynomial coordinates, not the transfer function, and improves numerical scaling.

In either representation, the sampled first variation can be written as
\[
 \delta Q_{i,k}=J_{N,i,k}D_Nu_N+J_{D,i,k}D_Du_D,
 \qquad \|u_N\|_2\le\tau_N,\quad \|u_D\|_2\le\tau_D.
\]
The $J$ matrices represent response derivatives, $u_N,u_D$ denote real decision vectors, and $D_N,D_D$ denote diagonal scaling matrices whose entries are the maximum of one and the corresponding parameter magnitude. The trust radii bound these scaled parameter increments.

\subsection{Error derivative and fitted phase directions}
\label{subsec:phase_template}
We use the directional derivative of the weighted error magnitude to construct the phase guidance. For $E_{i,k}=W_k(G_k-Q_{i,k})\ne0$,
\begin{equation}
 D_Q|E_{i,k}|[\delta Q]
 =\operatorname{Re}\{H_{i,k}\delta Q_{i,k}\},\qquad
 H_{i,k}=-\frac{\overline{E_{i,k}}W_k}{|E_{i,k}|}.
 \label{eq:weighted_magnitude_derivative}
\end{equation}
At one frequency, a first-order decreasing direction is $-\overline{H_{i,k}}$. Because a fixed-order rational increment couples the responses across frequencies, a common phase template is fitted at the selected peaks. Normalization is regularized near zero error, and negligible derivatives carry no reliable phase information.

\paragraph{Motivation from Hankel-norm approximation.}
We use Glover's all-pass construction \cite{Glover1984} to motivate the phase template. To relate this construction to the plant poles, let $G(s)=D_G+C_G(sI-A_G)^{-1}B_G$ be a minimal stable SISO realization with distinct poles. We use the eigenvalue decomposition
\begin{equation}
 A_G=V\Lambda V^{-1},\qquad
 \Lambda=\operatorname{diag}(p_1,\ldots,p_n),\qquad
 G(s)=D_G+\sum_{\ell=1}^{n}\frac{c_\ell}{s-p_\ell},
 \label{eq:plant_modal_eigendecomposition}
\end{equation}
where $p_\ell$ denotes a pole of $G$ and
$c_\ell=(C_Gv_\ell)(w_\ell^{\mathsf T}B_G)$ denotes its residue;
$v_\ell$ and $w_\ell^{\mathsf T}$ represent the corresponding column of $V$ and row of $V^{-1}$.

The approximation level is determined by a different eigenvalue problem. Let $X_c$ and $X_o$ denote the controllability and observability Gramians, respectively, satisfying
\[
 A_GX_c+X_cA_G^{\mathsf T}+B_GB_G^{\mathsf T}=0,\qquad
 A_G^{\mathsf T}X_o+X_oA_G+C_G^{\mathsf T}C_G=0.
\]
For a minimal stable realization, we use
\begin{equation}
 X_c^{1/2}X_oX_c^{1/2}
 =U\operatorname{diag}(\sigma_1^2,\ldots,\sigma_n^2)U^{\mathsf T},
 \qquad \gamma_H=\sigma_{m+1},
 \label{eq:hna_gramian_eigendecomposition}
\end{equation}
where $U$ is orthogonal and $\sigma_\ell$ denotes the $\ell$-th Hankel singular value, ordered nonincreasingly. Thus, \eqref{eq:plant_modal_eigendecomposition} identifies the interpolation locations, whereas \eqref{eq:hna_gramian_eigendecomposition} supplies the optimal order-$m$ Hankel-error level.

Glover's construction gives a stable approximant $Q_H$ and an antistable completion $R_H$ such that the completed error is scaled all-pass. In the scalar rational case, write this relation as
\begin{equation}
 G(s)-Q_H(s)-R_H(s)
 =\gamma_H\varepsilon_H\frac{d_E(-s)}{d_E(s)},\qquad
 \varepsilon_H\in\{-1,1\},
 \label{eq:hna_completed_allpass}
\end{equation}
where $d_E$ denotes the monic denominator of the completed error after cancellations. It contains the plant poles whenever $Q_H$ and $R_H$ are analytic there. If there are no cancellations and the denominators are pairwise coprime, $d_E=d_Gd_{Q_H}d_{R_H}$; the completion denominator has right-half-plane roots. The sign $\varepsilon_H\in\{-1,1\}$ determines the low-frequency orientation of the normalized all-pass factor.

The equality at a pole $p_\ell$ is understood after clearing denominators, since $G(p_\ell)$ is not finite. Assuming $Q_H$ and $R_H$ are analytic at $p_\ell$, matching residues in \eqref{eq:hna_completed_allpass} gives
\begin{equation}
 c_\ell=\gamma_H\varepsilon_H
 \frac{d_E(-p_\ell)}{d_E'(p_\ell)},
 \qquad \ell=1,\ldots,n.
 \label{eq:hna_pole_residue_interpolation}
\end{equation}
In the coprime case, the equivalent polynomial interpolation condition is
\begin{equation}
 n_G(p_\ell)=\gamma_H\varepsilon_H
 \frac{d_G(-p_\ell)d_{Q_H}(-p_\ell)d_{R_H}(-p_\ell)}
 {d_{Q_H}(p_\ell)d_{R_H}(p_\ell)}.
 \label{eq:hna_pole_polynomial_interpolation}
\end{equation}
These conditions are consequences of the complete all-pass identity; satisfying them alone does not determine a Hankel-optimal approximant. The completion is essential in the general construction, so $G-Q_H$ need not itself be all-pass.

For the special benchmarks in Appendix~\ref{app:benchmark_construction}, we instead use a stable approximation error of the form
\begin{equation}
 G(s)-Q_\star(s)=\gamma_\star\sigma_\star
 \frac{d_G(-s)d_{Q_\star}(-s)}{d_G(s)d_{Q_\star}(s)}.
 \label{eq:allpass_motivation}
\end{equation}
Here $G$ represents the reference transfer function in our approximation problem, and $\sigma_\star\in\{-1,1\}$ denotes the all-pass sign. The reflected denominator factors determine the phase while the error magnitude remains constant. We therefore retain the current approximation denominator in the phase template and use an adjustable reflected factor without requiring a realization of $G$. We fit this template to the magnitude-derivative phase, rather than imposing the Hankel interpolation equations on the FRD optimization.

We use the current Hurwitz denominator and an additional clustered factor to define the template:
\begin{equation}
 A_i(s;z)=\sigma_i\frac{D_i(-s)}{D_i(s)}
 \frac{\mathcal T_i(-s;z)}{\mathcal T_i(s;z)},\qquad
 \mathcal T_i(s;z)=(s+z)^r,\quad z>0.
 \label{eq:phase_template}
\end{equation}
Both denominator factors are Hurwitz, and $|A_i(j\omega;z)|=1$. Its phase is
\[
 \arg A_i(j\omega;z)=\arg\sigma_i-2\arg D_i(j\omega)
 -2r\arctan(\omega/z)\pmod{2\pi}.
\]
The sign $\sigma_i\in\{-1,1\}$ is updated from $\operatorname{sign}\operatorname{Re}(G-Q_i)$ at the smallest supplied frequency; the previous sign is retained when this real part is numerically zero. With $\widehat H_{i,k}=H_{i,k}/|H_{i,k}|$, the cluster location $z_i$ is selected by minimizing
\begin{equation}
 \frac{\sum_{k\in\mathcal A_i}v_{i,k}
 |\widehat H_{i,k}-A_i(j\omega_k;z)|^2}
 {\sum_{k\in\mathcal A_i}v_{i,k}},\qquad
 v_{i,k}\propto|H_{i,k}|^\nu.
 \label{eq:phase_fit}
\end{equation}
We use a coarse search in $\log z$, followed by local refinement within a positive frequency-scaled interval. The weights have a numerical floor, and $\nu=2$ by default. Since $|H_{i,k}|=|W_k|$ when the unregularized derivative is defined, these weights reflect the physical weight magnitude, rather than the error-peak height itself.

$v^Q_{i,k}=-\overline{A_i(j\omega_k;z_i)}$ represents the fitted response direction. The conjugation is essential: if the fit is exact and $\delta Q_{i,k}=a v^Q_{i,k}$ with $a>0$, then
\[
 D_Q|E_{i,k}|[\delta Q]=
 \operatorname{Re}\{H_{i,k}(-a\overline{A_i})\}=-a|H_{i,k}|<0.
\]
This first-order relation motivates the guidance constraints below; it does not certify a finite rational update.

\paragraph{Choice of template degree.}
The default $r=m+1$ makes the additional factor contribute an unwrapped phase change of $-(m+1)\pi$ from zero to infinite frequency. All roots remain at $-z$, so only one location is fitted. This degree and the peak-count limit are algorithmic defaults, not consequences of an alternation theorem. Other degrees can be specified for biproper and strictly proper approximants.

\subsection{Local minimax problem with soft phase guidance}
\label{subsec:weighted_socp}
For retained samples $k\in\mathcal I_i$, we use the linearized weighted residual $\widetilde E_{i,k}=E_{i,k}-W_k\delta Q_{i,k}$. Its magnitude is bounded by
\begin{equation}
 \left\|\begin{bmatrix}\operatorname{Re}\widetilde E_{i,k}\\
 \operatorname{Im}\widetilde E_{i,k}\end{bmatrix}\right\|_2
 \le\gamma_i+\delta\gamma.
 \label{eq:weighted_socp_cone}
\end{equation}
The right-hand side is constrained to be nonnegative, and its change is restricted relative to $\gamma_i$. Together with the parameter trust regions, these are second-order cone constraints \cite{Alizadeh2003}.

At each selected peak, we express the response increment relative to the fitted direction as
\[
 q_{i,k}=\overline{v^Q_{i,k}}\delta Q_{i,k},\qquad
 s_{i,k}=\max(|Q_{i,k}|,\epsilon_s).
\]
For an increment along $v^Q_{i,k}$, the rotated quantity $q_{i,k}$ is real and nonnegative. Departures from this direction are bounded by a common nonnegative slack $t_\phi$:
\begin{equation}
 \frac{|\operatorname{Im}q_{i,k}|}{s_{i,k}}\le t_\phi,
 \qquad \frac{\operatorname{Re}q_{i,k}}{s_{i,k}}\ge-\eta_\phi t_\phi,
 \qquad t_\phi\ge0,\quad k\in\mathcal A_i.
 \label{eq:phase_guidance_constraints}
\end{equation}
Here $\eta_\phi\ge0$ controls backward slack. At $t_\phi=0$, increments lie on the fitted nonnegative rays; positive slack permits departures. Thus $t_\phi$ bounds normalized response components, rather than an angular error.

The subproblem minimizes
\begin{equation}
 \min_{u_N,u_D,\delta\gamma,t_\phi}
 \frac{\delta\gamma}{\max(\gamma_i,\epsilon_\gamma)}+\lambda_i t_\phi
 \label{eq:guided_objective}
\end{equation}
subject to the cone, guidance, and trust-region constraints. The absolute-value constraint in \eqref{eq:phase_guidance_constraints} is represented by two linear inequalities. All response increments are affine in real variables, so the problem is an SOCP. Classical SCO is obtained by disabling the guidance constraints and penalty. If classical candidates are enabled, they are computed from the same iterate and compared with guided candidates after reconstruction and validation. 

\subsection{Trial validation and search adaptation}
\label{subsec:acceptance}
For each candidate increment, we use backtracking with $\alpha\in\{1,1/2,\ldots\,1/128\}$ to reconstruct the actual rational model. In pole--residue coordinates it updates the poles and residues; in coefficient coordinates it updates $N_i$ and $D_i$. Because reconstruction is nonlinear, the actual response generally differs from $Q_i+\alpha\delta Q_i$. Candidates with unstable poles, nonfinite responses, or invalid reconstruction are rejected; the local trust regions alone do not guarantee stability.

We screen admissible candidates and assess them on the complete input grid using
\begin{equation}
 M_\Omega(Q)=\gamma_\Omega(Q)+\pi(Q).
 \label{eq:weighted_merit}
\end{equation}
We use the optional pole-frequency regularizer
\[
 \pi(Q)=\kappa_\pi\exp\!\left[
 \min\{\max(\max_\ell|p_\ell|-\omega_c,0),L_\pi\}\right],
\]
where $\omega_c$ denotes a fixed cutoff and $L_\pi$ represents an upper limit on the exponent. This discourages very high pole frequencies. A varying penalty can trade error reduction against regularization, so merit decrease need not imply unpenalized error decrease. A trial is accepted only when its complete-grid merit decreases by the acceptance tolerance; otherwise the current model is retained.

Omitted samples that exceed the retained-grid bound are added to $\mathcal I_i$, and the subproblem is resolved. The complete grid is used after the permitted refinement passes if necessary. This checks the supplied samples, rather than introducing new frequency measurements.

Pole--residue candidates are tried first. If their best validated relative improvement is below $\tau_{\mathrm{switch}}=10^{-4}$, or modal conversion fails, coefficient candidates are computed from the same current model. The next iteration again starts with pole--residue coordinates. This threshold determines when to try alternative coordinates; it is distinct from the acceptance tolerance and the progress threshold used to adapt guidance and trust radii.

\paragraph{Optional frequency refinement.}
When $G$ and $W$ can be evaluated at arbitrary frequencies, the
validation grid is augmented after each iteration. Keeping the
resulting approximant $Q_{i+1}$ fixed, we refine the weighted-error
magnitude around each selected dominant local maximum, using its
neighbouring grid frequencies as an initial bracket, as explained in Section 2.7 of \cite{HitayOzbaybook}. Each local search evaluates a uniformly spaced frequency grid, retains the best point, and repeatedly narrows the bracket until the peak-magnitude change falls below a prescribed tolerance or the search limit is reached. Among the refined peaks, only the frequency $\omega_i^\star$ attaining the largest magnitude is retained:
\[
    \Omega_{i+1}=\Omega_i\cup\{\omega_i^\star\}.
\]
This sample is included in the next SOCP, and provides a better estimation of $\gamma_\infty$ rather than using initial samples only. Refinement does not change $Q_{i+1}$, trigger another SOCP solve, or reconsider its acceptance. The updated sampled error may therefore increase after grid augmentation. At most one frequency is added per iteration, until the prescribed sample limit is reached. Note that for FRD-only inputs, the supplied validation grid remains fixed.

\paragraph{Guidance and trust-region adaptation.}
The extremal phase guidance weighting coefficient $\lambda_i$ determines the relative importance of directional slack in \eqref{eq:guided_objective}. Larger values favour the fitted directions, whereas smaller values permit greater departures when they improve the predicted error bound. This coefficient is distinct from the physical weight $W$.

If no step is accepted, or validated relative progress is below \texttt{relTol}, the next guidance weight is
\begin{equation}
 \lambda_{i+1}=\max\{\lambda_{\min},\beta_\lambda\lambda_i\},
 \qquad 0<\beta_\lambda<1.
 \label{eq:extremal_weight_schedule}
\end{equation}
The defaults are $\lambda_0=1$, $\lambda_{\min}=10^{-4}$, and $\beta_\lambda=0.1$. Trust radii remain unchanged while guidance is reduced. They are contracted only when an iteration starts at $\lambda_{\min}$ and still provides insufficient progress. On adequate progress, guidance is retained and trust radii may increase up to their prescribed maxima. The positive $\lambda_{\min}$ leaves residual guidance, rather than recovering unguided SCO exactly. Termination uses the iteration limit or minimum trust radius.

\paragraph{Descent and its limits.}
Each accepted step decreases the merit on the current validation
grid. For a fixed grid and a nonnegative penalty, accepted merit
values form a nonincreasing sequence bounded below. Optional
frequency refinement changes the validation grid and can increase
the recorded merit without changing the approximant; hence this
fixed-grid descent argument does not establish convergence of the
sequence recorded across grid augmentations. Independent continuous-frequency verification remains necessary.

\begin{algorithm}[t]
\caption{Weighted EPGSCO for fixed-order model approximation}
\label{alg:weighted_epgsco}
\small

\noindent\textbf{Input:} State space realizations, symbolic representations, or FRD of $G,W$ on $\Omega$; stable initial model $Q_0$; order $m$; minimum relative degree $\rho$; algorithm settings.
\par\noindent\textbf{Output:} Stable approximant $Q$, sampled error $\gamma_\Omega(Q)$, and diagnostics.
\begin{enumerate}
\item Initialize the guidance weight and trust radii.
\item Repeat until the iteration limit or minimum trust radius is reached:\begin{enumerate}
\item Evaluate the complete-grid residual; select peaks $\mathcal A_i$ and optimization samples $\mathcal I_i$.
\item Fit \eqref{eq:phase_template} to the normalized derivative coefficients using \eqref{eq:phase_fit}.
\item Solve the guided pole--residue SOCP; also form classical candidates if enabled.
\item Backtrack and validate reconstructed candidates for stability and complete-grid merit decrease.
\item If pole--residue progress is below $\tau_{\mathrm{switch}}$ or modal conversion fails:\begin{enumerate}
\item Compute and validate scaled coefficient candidates from the same $Q_i$.
\end{enumerate}
\item If omitted samples violate the retained-grid bound:\begin{enumerate}
\item Add violating samples and resolve; use the complete grid if refinement is insufficient.
\end{enumerate}
\item Accept an admissible merit-decreasing candidate, if available; otherwise retain $Q_i$.
\item Adapt $\lambda_i$ and the trust radii according to validated progress.
\item Apply the optional frequency refinement if available, and
update the sampled error.
\end{enumerate}
\item Recompute $\gamma_\Omega(Q)$ on the final validation grid and return $Q$ and diagnostics.

\end{enumerate}
\end{algorithm}

\section{Numerical Examples}
\label{sec:numerical_examples}
The examples assess approximation accuracy, sensitivity to initialization, and computation time using unweighted and weighted CD-player reduction and two rational benchmarks with known optimal errors. The approximation methods use high-order rational surrogates of the original SISO systems obtained by vector fitting, while the reported errors are evaluated against the original systems. SCO and EPGSCO use identical settings for their common components, so their comparison assesses the combined contribution of phase guidance and adaptive frequency refinement.

For rational approximation, we distinguish between the sampled maximum
and the continuous-frequency norm:
\[
\gamma_{\mathrm{grid}}=\max_{\omega_k\in\Omega_{\mathrm{opt}}}
|W(j\omega_k)(P-P_a)(j\omega_k)|,
\qquad
\gamma_\infty=\sup_{\omega\in\mathbb R}|W(j\omega)(P-P_a)(j\omega)|.
\]
The definition of a sampled maximum does not itself certify the continuous-frequency norm. Independent verification uses the original plant and frequencies outside the optimization interval. The numerical values reported in the tables are $\gamma_\infty$.

We use YALMIP \cite{Lofberg2004} to formulate the local convex subproblems. HIFOO \cite{Burke2006} and MORLAB \cite{BennerSaakWerner2023MORLAB} represent the fixed-order optimization and model-reduction software, respectively. Table~\ref{tab:experimental_protocol} reports the software versions used. Table~\ref{tab:params_epgsco} presents the optimization parameters used in EPGSCO.

\begin{table}[h]\centering\small\color{black}
\caption{Experimental setup and software versions}
\label{tab:experimental_protocol}
\begin{tabularx}{\linewidth}{@{}p{0.55\linewidth}X@{}}\toprule
Item & Specification \\\midrule
Processor, RAM, operating system & i9-13900H, 32 GB, Windows \\
MATLAB & R2025b \\
HANSO & v2.02\\ 
HIFOO \cite{Burke2006} & v3.501 \\
YALMIP \cite{Lofberg2004} & v4.0 \\
MORLAB \cite{BennerSaakWerner2023MORLAB} & v6.0 \\
\bottomrule\end{tabularx}\end{table}
 
\begin{table}[htbp]\centering\small\color{black}
\caption{Parameters used in EPGSCO}
\label{tab:params_epgsco}
\begin{tabularx}{\linewidth}{@{}p{0.55\linewidth}X@{}}\toprule
Parameters & Used Values \\\midrule
SOCP solver & SDPT3\\
Solver and stopping tolerances & $10^{-5}$ \\
Pole--residue to polynomial switching tolerance & $10^{-4}$ \\
Number of starts & 1 \\
Iteration and time limits & 1000, \quad 1200 s \\
Extremal weight ($\lambda_i$),~$\beta_\lambda$ & $\lambda_0 = 1,~1\geq \lambda_i \geq 10^{-4}$,\quad $0.1$\\
$\tau_N,~\tau_D$ and their allowed interval & $\tau_{N,0}=0.2,\quad \tau_{D,0}=0.1,\quad 10^{-4}\leq \tau_{x,i}\leq 1^\dagger$\\ 
$\tau_g,~\tau_s$ & 2,\quad 0.25\\
Maximum samples used in SOCP & 4000 \\
\bottomrule\end{tabularx}
\par\smallskip
{\footnotesize
$\dagger$ The maximum trust-region radius is replaced by $10^3$ for $m>10$ to avoid early termination.}
\end{table}
\FloatBarrier
\subsection{CD player}
\label{subsec: ex1}
\begin{figure}[h]
    \centering
    \includegraphics[width=0.75\linewidth]{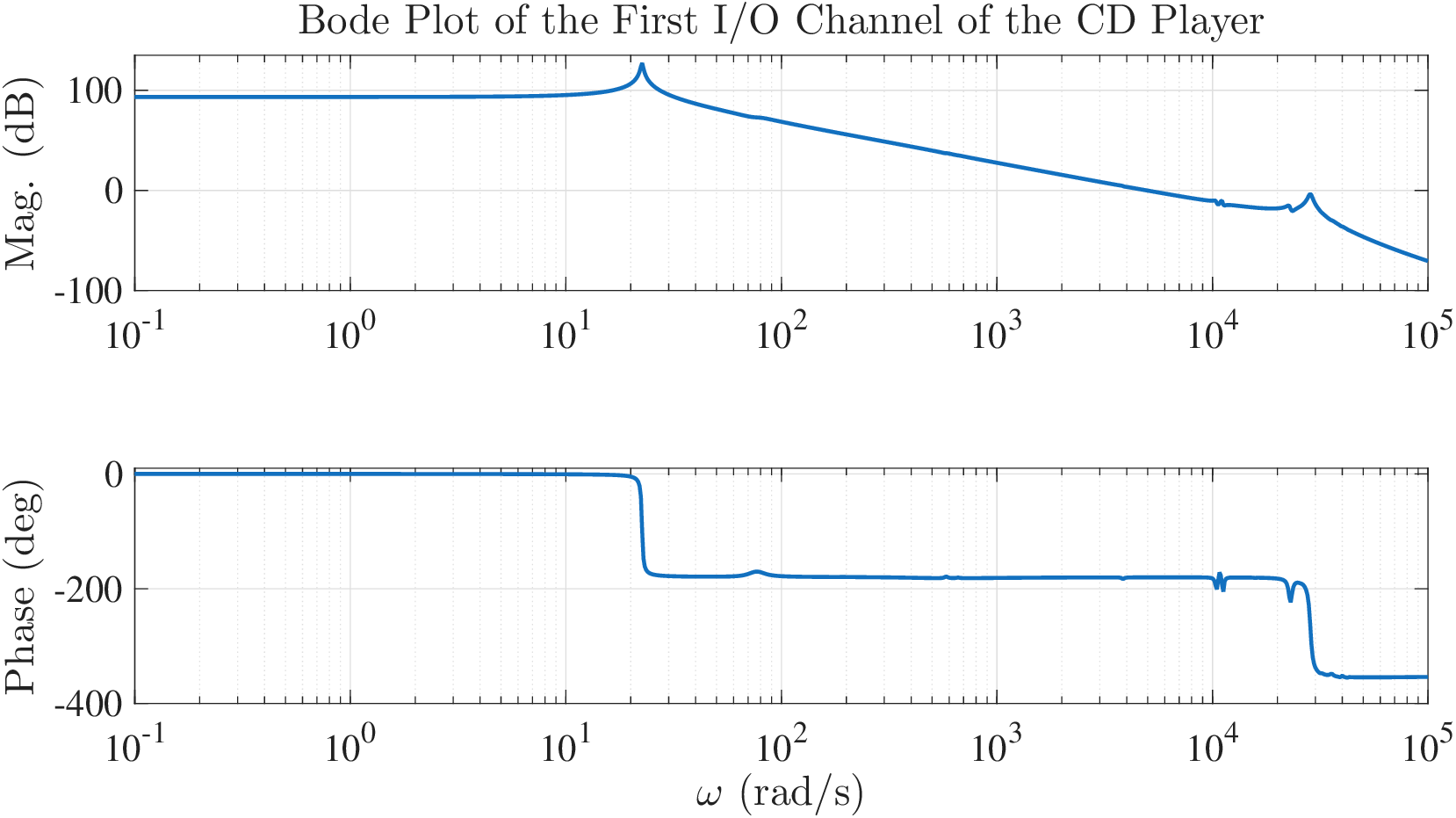}
    \caption{Bode plot of the first-input, first-output channel of the CD-player benchmark}
    \label{fig:ex1_bode_CD}
\end{figure}

We consider the CD-player benchmark from the SLICOT model-reduction collection \cite{Chahlaoui2002}. The original model $P$ has 120 states, two inputs, and two outputs. In this example, the subsystem from the first input to the first output is selected for both unweighted and frequency-weighted model-order reduction. All approximation methods use the order-49 rational surrogate $P_f$ obtained by vector fitting: FRD-based methods use its sampled
responses, while realization-based methods use its state-space representation. Optional frequency refinement is also applied based on the function $P_f$. Final approximation errors are evaluated against the original channel $P$. The $\Hi$-norm of the approximation error is $E_f = \norm{P-P_f}_\infty=0.0023567$. The Bode plot of the selected channel is shown in Fig.~\ref{fig:ex1_bode_CD}, using 1000 logarithmically spaced frequency samples over $\omega \in [0.1,10^5]$~rad/s. The same frequency set is used to sample $P_f$, which are used by the algorithms that operate on frequency-response data.

\subsubsection{Unweighted approximation}
\label{subsub:CD_unweighted}
The selected channel is approximated with $W=1$ for orders $m\in\{2,3,\ldots,10,13,15,20\}$. Note that the lower bound $\gamma_{lb}$ is found by
\[\gamma_{lb} =\max\left(\sigma_{f,(m+1)}- E_f,0\right)~,\]
where $\sigma_{f,(m+1)}$ represents the $(m+1)$-th Hankel singular value of $P_f$. 

\begin{table*}[th]
\centering
\caption{Approximation errors and computation times for the CD-player benchmark with $W=1$}
\vspace{0.3em}
\label{tab:cd_player_results}
\scriptsize
\setlength{\tabcolsep}{3pt}
\noindent\resizebox{\linewidth}{!}{%
\begin{tabular}{c|ccccc|ccccc|ccc}
\hline
& \multicolumn{5}{c|}{Baseline approximation errors}
& \multicolumn{5}{c|}{Optimization-based approximation errors}
& \multicolumn{3}{c}{Time (s)} \\
$m$
& $\gamma_{lb}$ & init. m. & $\gamma_{init}$ & $\gamma_{bt}$ & $\gamma_{hna}$
& $\gamma_{irka}$ & $\gamma_{\texttt{systune}}$ & $\gamma_{hifoo}$
& $\gamma_{sco}$ & $\gamma_{epgsco}$
& $t_{hifoo}$ & $t_{sco}$ & $t_{epgsco}$ \\
\hline
2  & 405.50 & \texttt{ssest} & 725.67 & 724.11 & 1028.2 & 724.11
   & 720.63 & 716.71 & 821.03 & 716.90 & 57.308 & 16.367 & 14.895 \\
3  & 327.76 & \texttt{ssest} & 644.05 & 655.48 & 334.30 & 724.12
   & 509.39 & 630.82 & 644.05 & 332.41 & 220.70 & 4.2808 & 66.001 \\
4  & 12.768 & \texttt{ssest} & 21.202 & 21.414 & 26.239 & 21.298
   & 16.342 & 20.842 & 17.556 & 16.192 & 46.350 & 13.699 & 42.363 \\
5  & 11.406 & \texttt{ssest} & 21.193 & 25.488 & 13.832 & 20.672
   & 12.877 & 20.591 & 18.924 & 11.564 & 27.477 & 7.7025 & 41.691 \\
6  & 4.3306 & \textit{vf} & 9.2847 & 6.8147 & 8.3302 & 7.0489
   & 4.8006 & 8.6572 & 6.8842 & 4.7918 & 238.15 & 7.6656 & 57.322 \\
7  & 3.8817 & \textit{vf} & 6.9970 & 6.5734 & 6.9641 & 7.6637
   & 4.2924 & 6.9335 & 4.6536 & 4.2862 & 35.689 & 5.3798 & 51.293 \\
8  & 2.1234 & \textit{vf} & 3.2212 & 3.4978 & 4.5592 & 3.5315
   & 3.2218 & 3.2097 & 2.9407 & 2.8609 & 35.276 & 3.9148 & 83.594 \\
9  & 1.8323 & \textit{vf} & 2.8279 & 3.2218 & 3.1537 & 3.4110
   & 2.4966 & 2.8271 & 3.2498 & 2.4647 & 52.088 & 7.9366 & 45.147 \\
10 & 1.5508 & \textit{vf} & 2.8809 & 2.7961 & 2.3452 & 3.0682
   & 1.6978 & 2.8782 & 3.4752 & 1.6639 & 5.4272 & 3.9811 & 92.232 \\
13 & 0.58681 & \texttt{ssest} & 1.4061 & 1.2610 & 1.0029 & --
   & 0.71597 & 1.3985 & 0.86705 & 0.70815 & 7.5289 & 8.2154 & 200.76 \\
15 & 0.43269 & \texttt{ssest} & 0.96571 & 0.82574 & 0.87363 & --
   & 0.65739 & 0.96146 & 0.72002 & 0.49112 & 10.635 & 19.132 & 158.78 \\
20 & 0.087221 & \texttt{ssest} & 0.55985 & 0.17074 & 0.27780 & 0.31784
   & 0.21630 & 0.55985 & 0.11863 & 0.10248 & 56.338 & 7.1011 & 134.69 \\
\hline
\end{tabular}%
}
\end{table*}

The $\Hi$-norm of approximation errors $\gamma_x = \norm{P-P_{a,x}}_\infty$, where $x$ denotes the chosen method and $P_{a,x}$ is the corresponding approximation, and the computation times for the compared optimization methods are reported in Table~\ref{tab:cd_player_results}.

\begin{figure}[htbp]
  \centering
  \begin{minipage}{0.48\textwidth}
    \centering
    \includegraphics[width=\linewidth]{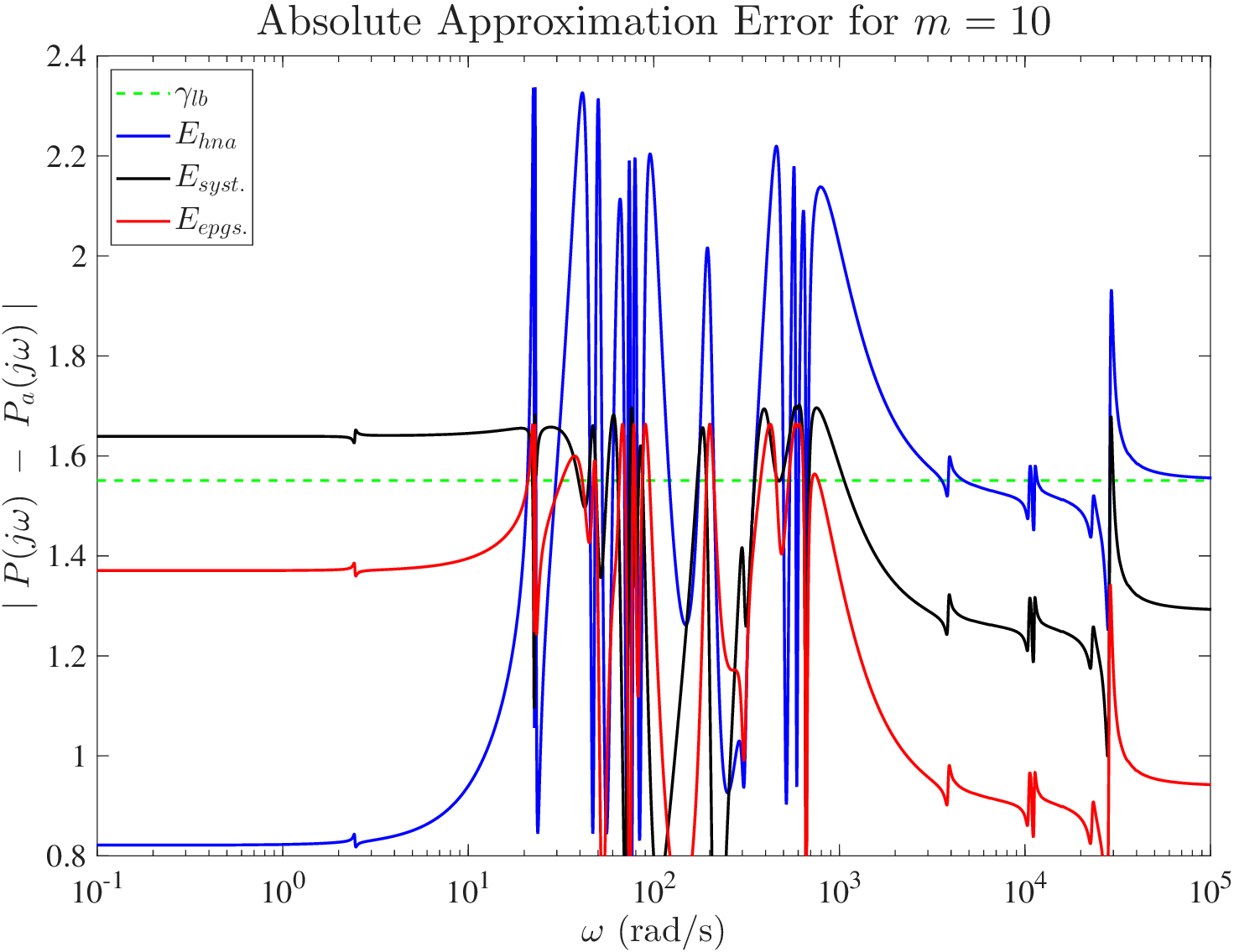}
    \caption{Absolute error for $W=1,~m=10$}
    \label{fig:ex1_uw_m10}
  \end{minipage}
  \hfill
  \begin{minipage}{0.48\textwidth}
    \centering
    \includegraphics[width=\linewidth]{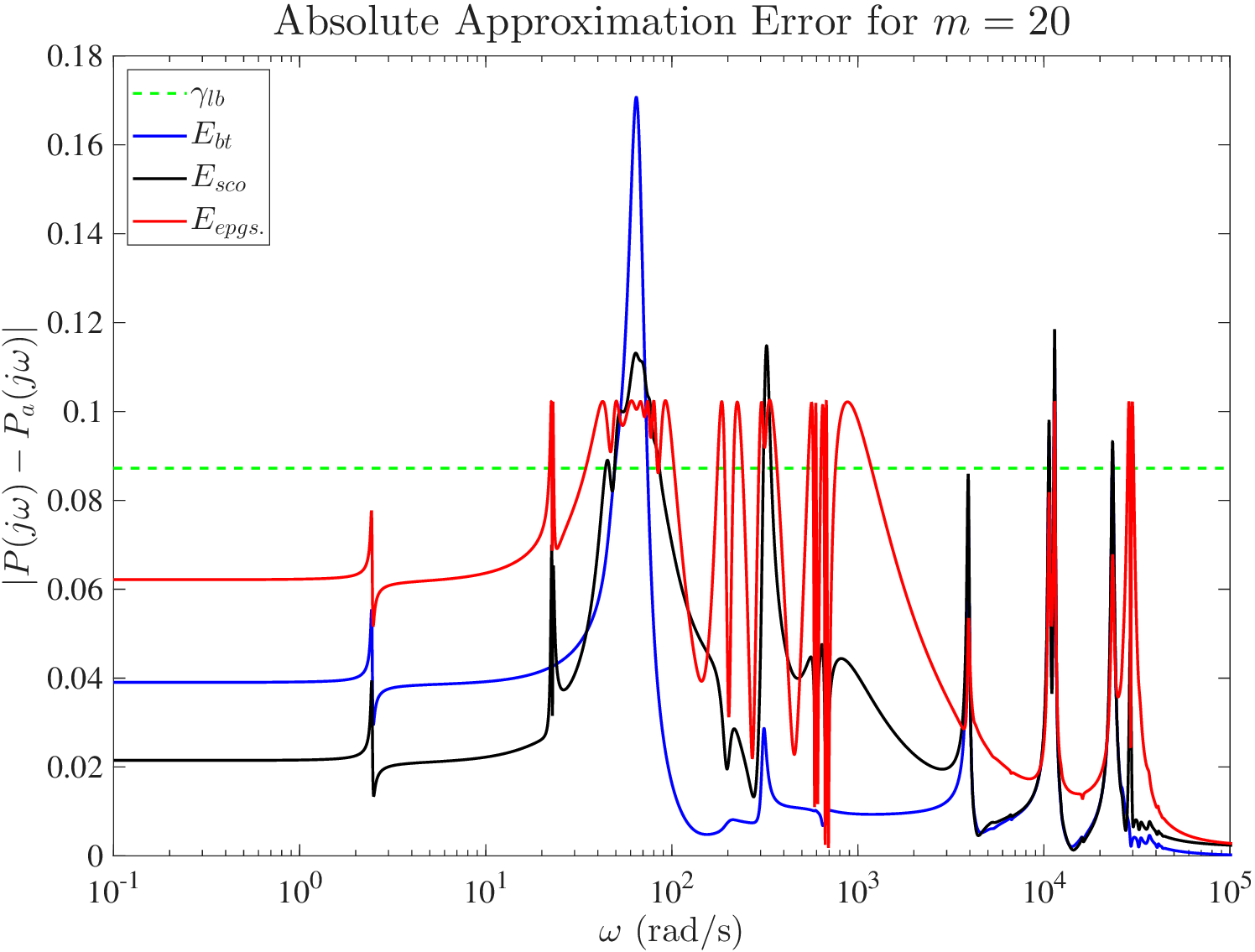}
    \caption{Absolute error for $W=1$, $m=20$.}
    \label{fig:ex1_uw_m20}
  \end{minipage}
\end{figure}

The initial approximations are obtained using \texttt{ssest} and vector fitting ($vf$). For each order $m$, the approximation with the smaller $\Hi$-norm error is selected. At $m=13$ and $m=15$, IRKA fails to converge to a stable approximation within 1000 iterations. At $m=20$, EPGSCO reduces the reported error from $0.11863$ for SCO to $0.10248$, corresponding to an improvement of approximately $13.6\%$. At $m=2$, HIFOO achieves a slightly smaller approximation error than EPGSCO ($716.71$ versus $716.90$). At all other tested orders, EPGSCO achieves the smallest approximation error among the compared methods. Figs.~\ref{fig:ex1_uw_m10} and~\ref{fig:ex1_uw_m20} show the approximation-error magnitudes for $m=10$ and $m=20$, respectively, evaluated over the same frequency interval using a denser grid of $10^4$ samples than that used by the algorithms.

\subsubsection{Weighted approximation}
The following weight is introduced to emphasize approximation accuracy near the resonance frequencies $\omega_1$ and $\omega_2$,
\[
    W(s) = 1+9\left(\frac{2\zeta \omega_1 s}{s^2 + 2\zeta \omega_1 s + \omega_1^2}\right) + 9\left(\frac{2\zeta \omega_2 s}{s^2 + 2\zeta \omega_2 s + \omega_2^2}\right)~,
\]
where $\omega_1=22.565$, $\omega_2=28500$ rad/s, and $\zeta=0.1$ is the damping ratio. This choice of weight amplifies the approximation error by approximately a factor of ten near each resonance frequency $\omega_i$, $i=1,2$.

As in Section~\ref{subsub:CD_unweighted}, 1000 logarithmically spaced samples over $[0.1,10^5]$~rad/s are selected initially for optimization, while the same $P_f$ is used for state space representation based methods. Final $\Hi$ approximation errors are calculated based on the original channel with 120 states, and shown in Table~\ref{tab:cd_player_weighted_results}.

For the calculation of the lower bound for the weighted $\Hi$ approximation error, denoted by $\gamma_{wlb}$, the following triangle inequality is used. Let $G_{\mathrm{fit}}$ be a stable fitted model and let $L_m(G_{\mathrm{fit}},W)$ be a valid order-$m$ lower bound for the same weighted approximation problem, which can be calculated by using the bound given in \cite{LathamAnderson1985}. If
\[
\|W(P-G_{\mathrm{fit}})\|_\infty\leq\varepsilon_{\mathrm{fit}},
\]
the triangle inequality gives the corrected bound
\begin{equation}
\gamma_{{wlb},m}
=\max\{L_m(G_{\mathrm{fit}},W)-\varepsilon_{\mathrm{fit}},0\}.
\label{eq:corrected_fitting_lower_bound}
\end{equation}

EPGSCO achieves the smallest weighted approximation error among the compared methods at all tested orders. At $m=20$, the reported error decreases from $0.69752$ for SCO to $0.15575$ for EPGSCO, corresponding to a reduction of approximately $77.7\%$. The EPGSCO error is approximately $6.7\%$ above the displayed lower bound of $0.14598$. For $m=5$ and $m=15$, IRKA fails to return a stable approximation within 1000 iterations.

\begin{table*}[t]
\centering
\caption{Weighted approximation errors and computation times for the CD-player benchmark}
\vspace{0.3em}
\label{tab:cd_player_weighted_results}
\scriptsize
\setlength{\tabcolsep}{3pt}
\noindent\resizebox{\linewidth}{!}{%
\begin{tabular}{c|ccc|ccccc|ccc}
\hline
& \multicolumn{3}{c|}{Baseline approximation errors}
& \multicolumn{5}{c|}{Optimization-based approximation errors}
& \multicolumn{3}{c}{Time (s)} \\
$m$
& $\gamma_{wlb}$ & $\gamma_{init}$ & $\gamma_{wbt}$
& $\gamma_{irka}$ & $\gamma_{\texttt{systune}}$ & $\gamma_{hifoo}$
& $\gamma_{sco}$ & $\gamma_{epgsco}$
& $t_{hifoo}$ & $t_{sco}$ & $t_{epgsco}$ \\
\hline
2  & 500.45 & 867.60 & 1559.9
   & 1710.3 & 862.03 & 867.60 & 922.71 & 861.28 & 12.435 & 6.2701 & 32.253 \\
3  & 406.97 & 798.14 & 1471.9
   & 1709.0 & 687.89 & 798.14 & 785.24 & 685.50 & 63.471 & 11.754 & 27.579 \\
4  & 19.886 & 35.066 & 32.813
   & 35.845 & 29.950 & 35.066 & 27.381 & 26.208 & 52.189 & 15.956 & 22.175 \\
5  & 19.211 & 34.466 & 33.467
   & -- & 36.303 & 34.466 & 29.211 & 24.129 & 21.415 & 6.0215 & 31.261 \\
6  & 5.7559 & 13.488 & 13.142
   & 66.054 & 11.273 & 13.488 & 9.6941 & 8.5941 & 56.817 & 12.215 & 35.631 \\
7  & 4.6847 & 8.9466 & 14.834
   & 47.663 & 7.2131 & 8.9466 & 8.0447 & 6.5245 & 87.210 & 7.0792 & 77.214 \\
8  & 2.9534 & 6.9392 & 6.7211
   & 50.258 & 6.9392 & 7.0850 & 6.8543 & 6.6079 & 45.018 & 15.181 & 377.55 \\
9  & 2.8392 & 6.6172 & 21.555
   & 30.874 & 6.6172 & 6.6167 & 7.4088 & 3.2759 & 91.301 & 20.031 & 382.47 \\
10 & 2.1774 & 3.6736 & 4.6779
   & 50.104 & 3.7324 & 3.6736 & 3.6736 & 2.9869 & 60.630 & 4.5733 & 133.03 \\
13 & 1.1720 & 2.8769 & 4.9418
   & 37.219 & 2.7005 & 2.8769 & 3.4514 & 1.4611 & 41.404 & 9.4708 & 126.46 \\
15 & 0.59986 & 2.4589 & 1.3586
   & -- & 2.2073 & 2.4579 & 0.97838 & 0.77543 & 20.296 & 15.732 & 209.26 \\
20 & 0.14598 & 0.91403 & 0.26825
   & 35.982 & 1.0967 & 1.0388 & 0.69752 & 0.15575 & 118.11 & 11.575 & 899.68 \\
\hline
\end{tabular}%
}
\end{table*}

\FloatBarrier
\subsection{Plant with Known Optimal $\Hi$ Approximation}
Two constructed plants with known optimal $\Hi$ approximation errors (Appendix~\ref{app:benchmark_construction}) are used to assess attainment of a certified optimum and dependence on initialization. The examples contain lightly damped modes and distinct plant and approximant pole configurations. The tenth-order plant is reduced to order four, and the $42^{nd}$-order plant to order eight. The optimization inputs are frequency-response samples. For implementations requiring state-space models, vector-fitting surrogates are used, which have the same order as the introduced plants. The exact plants are retained for the calculation of resulting $\Hi$-norms of the approximation errors.

\subsubsection{Tenth Order Plant with Fourth Order Optimal $\Hi$ Approximation}
\label{subsec: ex2}
Let a tenth order stable system $P$ be given as
\newline

\resizebox{0.95\linewidth}{!}{$
P_{10}(s)=
\frac{
1.5 (s-52.55) (s-0.09643) (s^2 - 0.04582s + 0.132) 
(s^2 - 0.9265s + 8.35)(s^2 - 1.039s + 167.8) (s^2 - 53.36s + 9322)
}{
(s+200) (s+0.1) (s^2 + 0.3s + 0.25) 
(s^2 + 1.4s + 4) (s^2 + 3s + 225) (s^2 + 50s + 2500)
}~.
$}
\newline

Note that $P_{10}(s)$ is an equal-tail HSV system at reduction order \(4\), with
\[
    \sigma_4(P_{10})=1.545>\sigma_5(P_{10})=\ldots=\sigma_{10}(P_{10})=1
\]
The fourth order $\Hi$ optimal approximation is 
\[
    P_{opt,4}(s) = \frac{0.5 (s-34.47) (s-0.08938) (s^2 - 2.188s + 42.84)}{(s^2 + 0.5s + 1) (s^2 + 6s + 400)}~.
\]

In this problem setting, given 1000 frequency response samples of $P(j\omega)$, with $\omega$ being logarithmically separated points in the interval $\left[10^{-1},~10^4\right]$ \(rad/s\), we are asked to find a $4^{th}$ order $\Hi$ approximation by using the same methods used in Sec.~\ref{subsec: ex1}. The initial approximation candidate for the algorithms is obtained via vector fitting of the frequency response data, which is obtained as
\[
    P_{vf,4}(s) = \frac{0.40551 (s-5.527) (s-0.0764) (s^2 - 0.07577s + 0.1187)}{(s^2 + 0.2487s + 0.06911) (s^2 + 0.6654s + 0.6471)}~.
\]

\begin{figure}[h]
    \centering
    \includegraphics[width=0.8\linewidth]{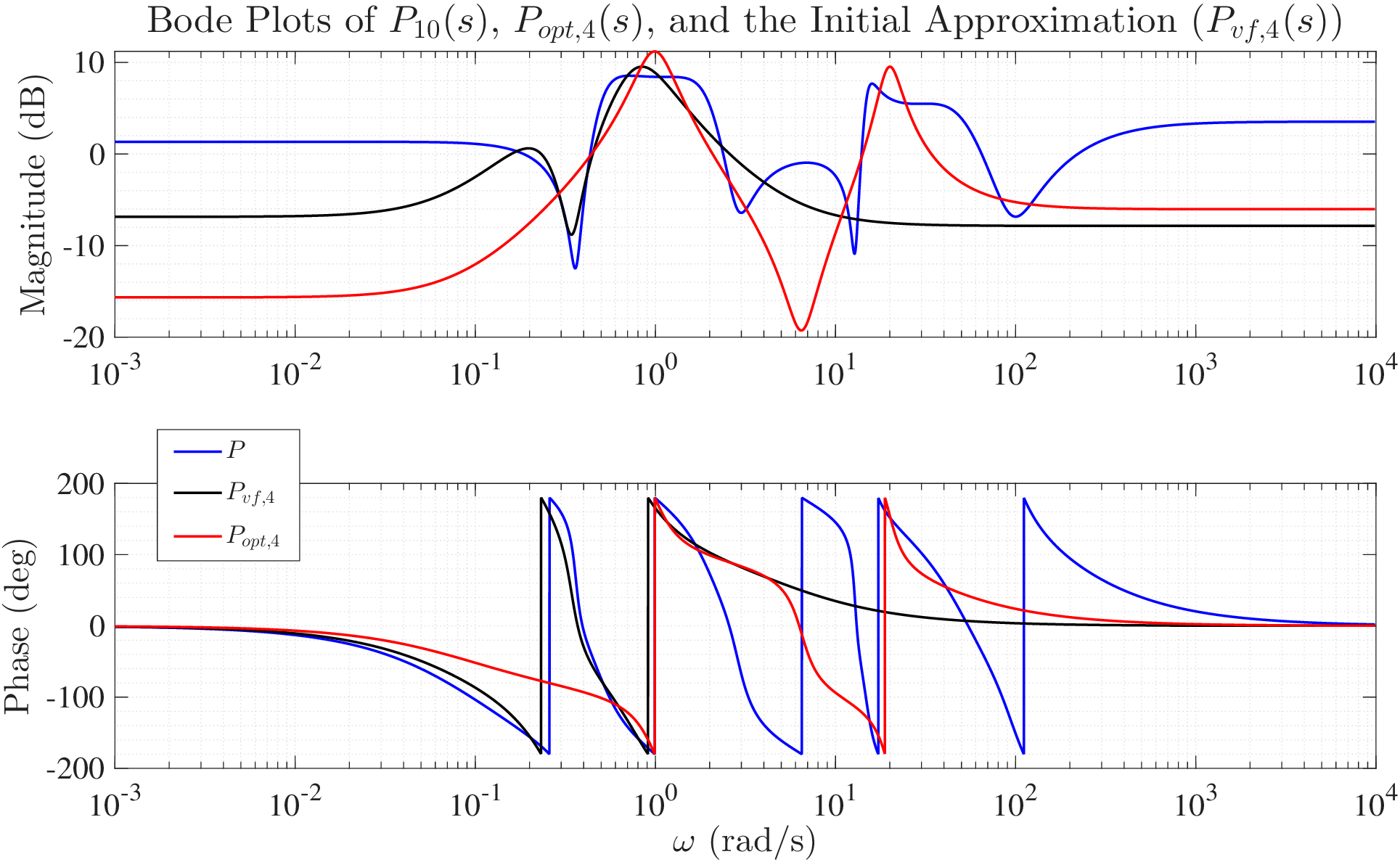}
    \caption{Bode plots of the 10-th order plant, $P_{opt,4}(s)$, and the initial approximation $P_{vf,4}(s)$}
    \label{fig:ex2_bode_P10}
\end{figure}

The Bode plots of the plant $P_{10}(s)$, $\Hi$ optimal approximation $P_{opt,4}(s)$, and the initial approximation candidate are also shown in Fig.~\ref{fig:ex2_bode_P10} to illustrate the plant dynamics and difference between initial and the optimal approximations in terms of their modes.

\begin{table}[th]
\centering
\caption{Approximation errors and computation times for $\Hi$ approximation of $P_{10}(s)$.}
\vspace{0.2em}
\label{tab:ex2_P10_results}
\scriptsize
\setlength{\tabcolsep}{3pt}
\noindent\resizebox{\linewidth}{!}{%
\begin{tabular}{c|ccc|ccccc|cccc}
\hline
& \multicolumn{3}{c|}{Baseline approximation errors}
& \multicolumn{5}{c|}{Optimization-based approximation errors}
& \multicolumn{4}{c}{Time (s)} \\
$m$
& $\sigma_{m+1}$
& Initial ($vf$)
& BT
& IRKA
& HIFOO
& \texttt{systune}
& SCO
& EPGSCO
& HIFOO
& \texttt{systune}
& SCO
& EPGSCO \\
\hline
4
& 1.000
& 2.846
& 2.000
& 2.644
& 1.514
& 2.072
& 1.912
& 1.000
& 50.78
& 1.108
& 32.52
& 58.26 \\
\hline
\end{tabular}%
}
\end{table}
\begin{figure}[H]
    \centering
    \includegraphics[width=0.7\linewidth]{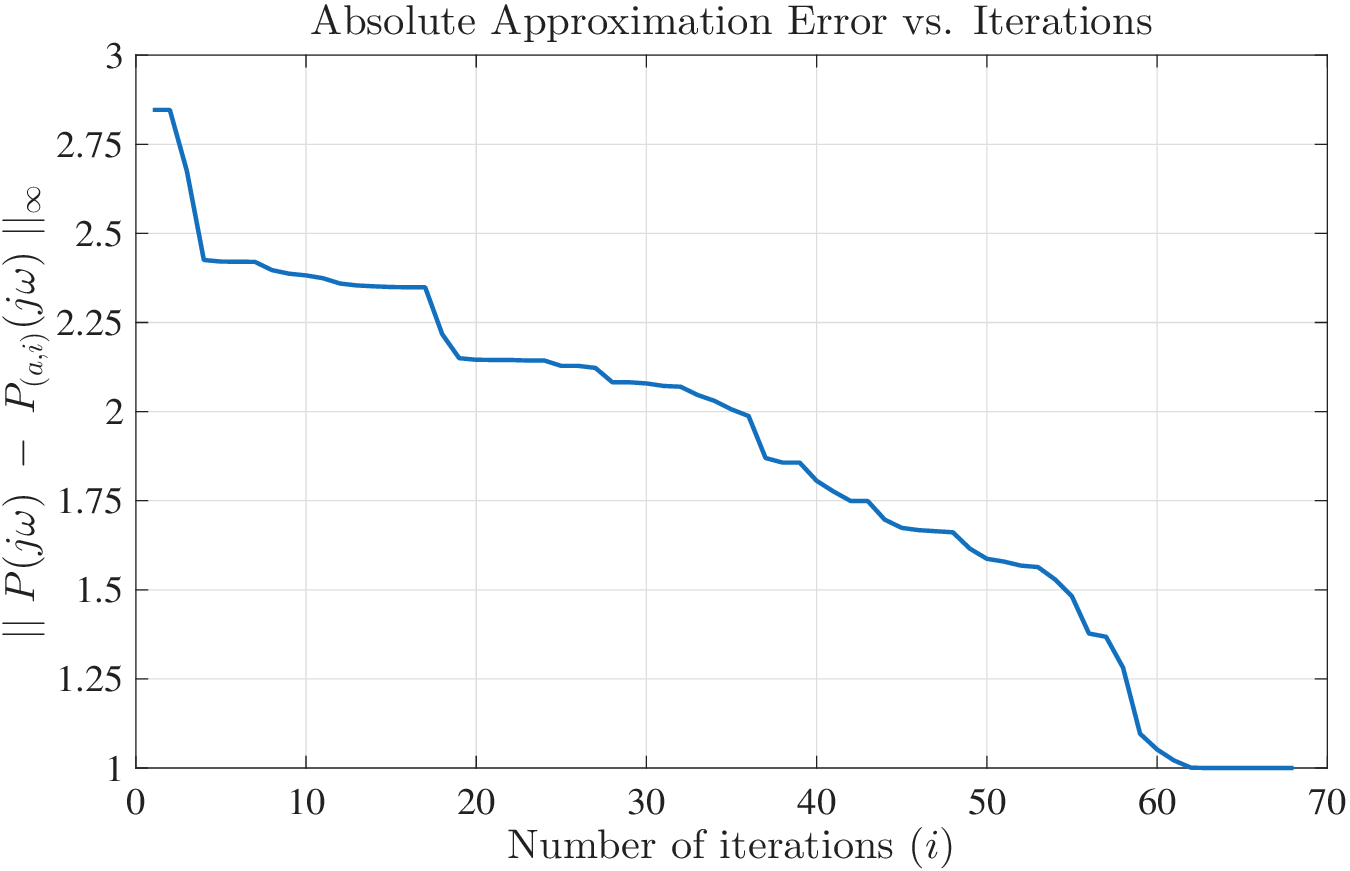}
    \caption{Maximum absolute error through iterations in EPGSCO for $P_{10}(s)$}
    \label{fig:ex2_P10_gamma}
\end{figure}
The decay of maximum absolute error with respect to iterations, denoted by $\gamma_{epgsco,i}$, is shown in Fig.~\ref{fig:ex2_P10_gamma}. The displayed error reaches $1.0000$ at iteration $62$, and the search terminates at iteration $68$ when the trust-region radius reaches $10^{-4}$.

\subsubsection{Moderate Order Plant with an Eighth Order Optimal $\Hi$ Approximation}

\begin{figure}[h]
    \centering
    \includegraphics[width=0.8\linewidth]{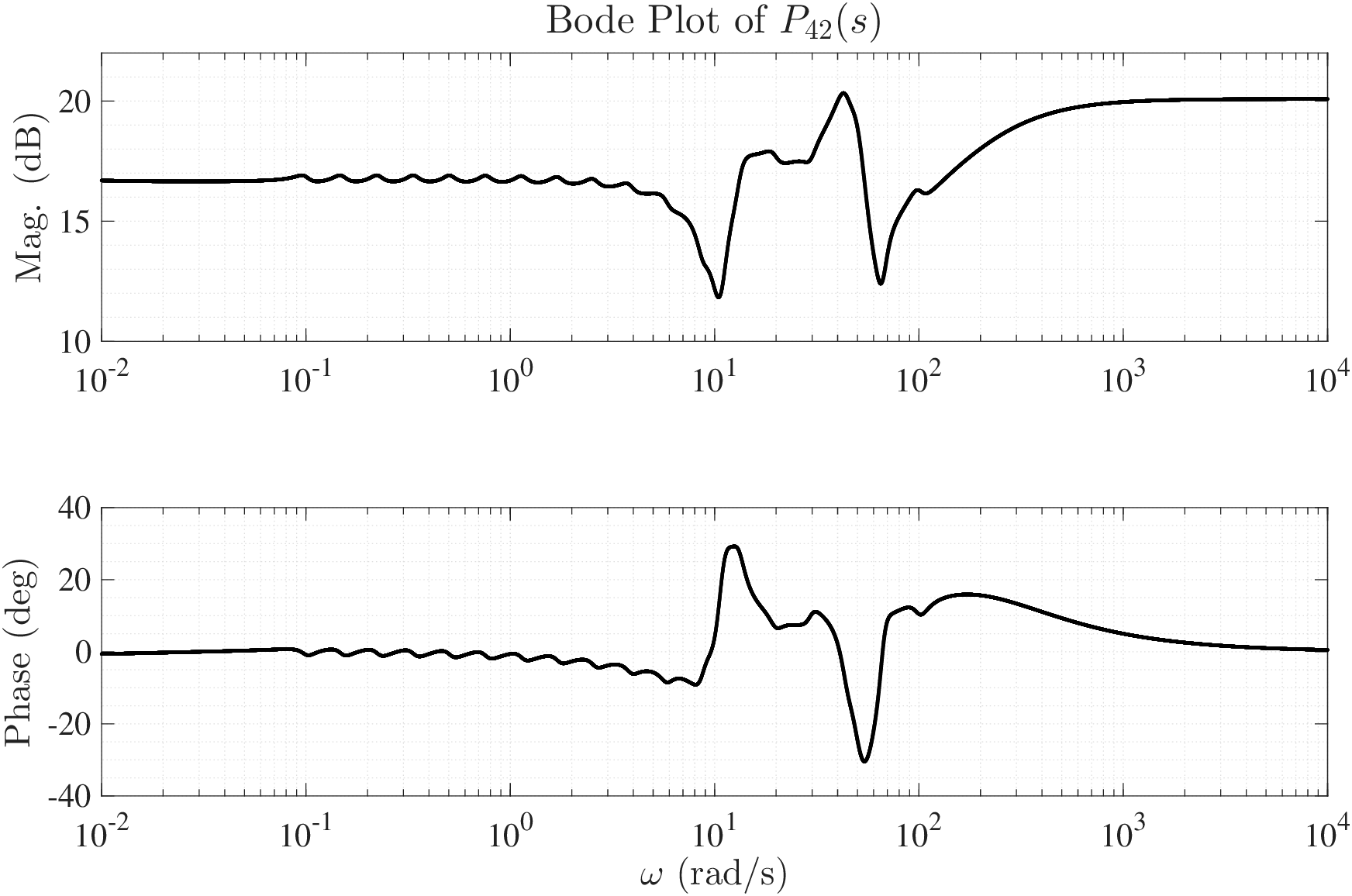}
    \caption{Bode plot of $P_{42}(s)$}
    \label{fig:ex2_P42_bode}
\end{figure}

A $42^{nd}$ order plant, $P_{42}(s)$, with a known eighth order
optimal $\Hi$ approximation $P_{\mathrm{opt},8}(s)$ is constructed to compare approximation algorithms on a moderate order function. Hankel singular values of the plant satisfy
\[
\sigma_9(P_{42})=\cdots=\sigma_{42}(P_{42})=0.1,
\]
and the optimal approximation error is $0.1$. The denominator of
$P_{\mathrm{opt},8}$ is prescribed as
\[
d_{\mathrm{opt},8}(s)
=\prod_{i=1}^{4}
\left(s^2+2\zeta_{a,i}\omega_{a,i}s+\omega_{a,i}^2\right),
\]
where
\[
(\zeta_{a,i})_{i=1}^{4}=(0.25,\,0.35,\,0.4,\,0.9),
\qquad
(\omega_{a,i})_{i=1}^{4}=(12,\,45,\,60,\,90).
\]
The plant has real poles at $-0.005$ and $-200$, together with
20 complex conjugate pole pairs
\[
p_{i,\pm}
=-\zeta_{P,i}\omega_{P,i}
\pm \mathrm{j}\omega_{P,i}\sqrt{1-\zeta_{P,i}^{2}},
\qquad i=1,\ldots,20.
\]
For $i=1,\ldots,18$, the natural frequencies $\omega_{P,i}$ are
logarithmically spaced from $10^{-1}$ to $10^2$, with
$\zeta_{P,i}=0.1$. The remaining two pairs have
$(\omega_{P,19},\omega_{P,20})=(11,50)$ and
$\zeta_{P,19}=\zeta_{P,20}=0.15$. In addition, the high-frequency gains are set to $P_{\mathrm{opt},8}(\infty)=10$ and $P_{42}(\infty)=10.1$. The remaining parameters of both transfer functions are determined from the interpolation conditions in \eqref{eq:numerator_interpolation} and the gain condition in \eqref{eq:high_frequency_gain_condition}. Fig.~\ref{fig:ex2_P42_bode} shows the Bode plot of this plant, $P_{42}(s)$.

\begin{table*}[h]
\centering
\caption{Approximation errors and computation times for $\Hi$ approximation of $P_{42}(s)$ with different initializations and $m=8$.}
\vspace{0.3em}
\label{tab:ex2_P40_results}

\scriptsize
\setlength{\tabcolsep}{4pt}
\noindent\resizebox{\linewidth}{!}{%
\begin{tabular}{c|cc|ccccc|ccc}
\hline
& \multicolumn{2}{c|}{Baseline errors}
& \multicolumn{5}{c|}{Optimization-based approximation errors}
& \multicolumn{3}{c}{Time (s)} \\
Method
& $\sigma_{m+1}$
& Initial
& IRKA
& HIFOO
& \texttt{systune}
& SCO
& EPGSCO
& HIFOO
& SCO
& EPGSCO \\
\hline
BT
& \multirow{4}{*}{0.1000} & 0.2000 & 0.1871 & 0.1734 & 0.1000
& 0.1000 & 0.1000 & 430.51 & 16.608 & 126.53 \\
\texttt{fitfrd}
& & 3.5184 & 0.1890 & 1.4988 & 1.2957
& 100.21 & 0.1000 & 425.34 & 59.362 & 248.16 \\
\texttt{ssest}
& & 10.100 & 0.1877 & 0.8229 & 0.7130
& 10.100 & 0.1000 & 430.19 & 3.3232 & 215.09 \\
VF
& & 1.7327 & 0.1904 & 0.1000 & 0.5527
& 48868 & 0.1000 & 49.803 & 41.382 & 200.98 \\
\hline
\end{tabular}%
}
\end{table*}

For the \texttt{fitfrd} and VF initializations, the SCO approximants exhibit large continuous-frequency errors despite reductions in the sampled objective. This discrepancy highlights the importance of independent verification between and beyond the optimization samples. For the \texttt{fitfrd} initialization, SCO reduces the maximum sampled error to $\gamma_\Omega(Q_{\mathrm{SCO}})=0.2941$, whereas the reported continuous-frequency error is $100.21$, as shown in Table~\ref{tab:ex2_P40_results}. This significant mismatch is associated with a near-origin pole--zero pair, $z_1\approx-1.5\times10^{-7}$ and $p_1=-10^{-8}$, whose low-frequency behavior is not captured by the supplied FRD samples. In contrast, EPGSCO attains the known globally optimal error of $0.1000$, to the reported precision, from all four tested initializations.

\FloatBarrier
\section{Conclusion}
In this paper, we proposed an extremal-phase-guided sequential convex
optimization algorithm for stable fixed-order frequency-weighted
$\mathcal H_\infty$ approximation. The phase guidance is motivated by
the all-pass structure used in Glover's Hankel-norm approximation and
incorporated into local SOCP updates of the poles and residues. We use
dominant-error-peak selection and focused frequency grids to construct
the optimization subproblems, while scaled polynomial-coefficient
updates are considered when the pole--residue step provides
insufficient improvement. Regularization, trust regions, and
backtracking govern the trial models. The extremal weight, which
denotes the coefficient of the phase-guidance penalty, is decreased
when necessary to allow greater departures from the fitted directions,
while stability checks and nonlinear complete-grid validation
determine whether a candidate is accepted. When arbitrary-frequency
evaluation is available, optional peak refinement adds at most one
frequency per iteration; with sufficient initial frequency coverage,
the refined $\gamma_{\mathrm{grid}}$ can closely estimate
$\gamma_\infty$, subject to independent verification.

We evaluated the proposed algorithm using unweighted and weighted
CD-player reduction and rational benchmarks with known optimal
approximation errors and compared the results with state-of-the-art
methods. The reported results show that EPGSCO can improve the accuracy
obtained by classical SCO and the other compared methods, although
additional computation is required in some cases. For the constructed
$P_{42}$ system, EPGSCO attains the known optimal error from all four
tested initializations, whereas several compared methods return
suboptimal approximations for some initializations. These results
demonstrate the benefit of the proposed algorithm and also indicate
that its accuracy and computation time depend on the problem,
approximation order, and initialization. The acceptance rule provides
decreasing merit values on the current grid, although subsequent
frequency refinement may increase the recorded sampled error;
a general global-optimality result has not yet been established.

As future work, we are investigating the optimality conditions for
fixed-order $\mathcal H_\infty$ approximation, the lowest achievable
approximation error for a given order, and the construction of an
optimal $\mathcal H_\infty$ approximant. In particular, we aim to
determine how the all-pass phase structure, extremal frequencies, and
rational interpolation conditions can characterize the optimal error
and distinguish a locally improving approximation from an optimal
one. For weighted and structured problems, this requires accounting
for the weight and the admissible approximation class. Further work
includes extending the proposed algorithm to fixed-order
$\mathcal H_\infty$ controller design.

\section*{Acknowledgment}
The presented research was supported by the Czech Science Foundation under the project 24-10301S. The first author was supported by Postdoctoral Fellowship Programme MSCA COFUND-CZ CROP  CZ.02.01.01/00/22\_011/0008569.

\appendix
\section{Known-Optimum Benchmark Construction}
\label{app:benchmark_construction}
This appendix collects the construction used in Section~\ref{sec:numerical_examples}. It supplies a benchmark certificate under its stated assumptions, rather than an optimality theorem for the phase-guided algorithm. The equal-tail result allows a freely chosen feedthrough term and does not impose strict properness on the approximant.

\begin{definition}[Equal-tail HSV system]
A stable minimal system \(P\) of McMillan degree \(n\) is called an
\emph{equal-tail Hankel-singular-value (HSV) system at reduction
order \(m\)}, where \(1\leq m<n\), if its Hankel singular values satisfy
\[
    \sigma_m(P)>
    \sigma_{m+1}(P)=\cdots=\sigma_n(P)>0~.
\]
\end{definition}
\begin{proposition}[$\Hi$ optimality for an equal HSV tail]
Let $P\in\RHinf$ be a minimal SISO transfer function of
McMillan degree $n$, and let $1\leq m<n$. Assume that $P$ is an equal-tail HSV system at reduction order $m$. Then there exists a stable transfer function $P_{a,m}$ of McMillan degree at most $m$ such that
\begin{equation}
\|P-P_{a,m}\|_\infty=\sigma_{m+1}(P).
\label{eq:equal_tail_optimal_error}
\end{equation}
Consequently,
\[
\inf_{\substack{Q\in\RHinf\\ \deg Q\leq m}}
\|P-Q\|_\infty
=
\sigma_{m+1}(P).
\]
\end{proposition}

\begin{proof}
For every stable $Q$ of McMillan degree at most $m$,
the Hankel singular value bound yields
\[
\sigma_{m+1}(P)\leq\|P-Q\|_\infty.
\]
On the other hand, Glover's $\Hi$ error bound \cite{Glover1984} for a
suitably selected optimal Hankel-norm approximation
sums the distinct HSV levels in the discarded tail,
counting each level only once. Since the discarded
tail contains only the level $\sigma:=\sigma_{m+1}(P)$, this result gives
a stable approximant $P_{a,m}$ of degree at most $m$
for which
\[
\|P-P_{a,m}\|_\infty\leq\sigma.
\]
The lower and upper bounds coincide, proving
\eqref{eq:equal_tail_optimal_error}.
\end{proof}

Given the denominators of a plant and an order-$m$ approximant, together
with a target error level $\gamma$, their numerators can be constructed
so that the approximation error is all-pass. The following proposition
gives the corresponding interpolation conditions.
\begin{proposition}[Construction of the numerator polynomials]
Let $d_P$ and $d_{P_a}$ be real monic Hurwitz polynomials of degrees
$n$ and $m$, respectively, whose roots
$\{p_i\}_{i=1}^{n}$ and $\{q_j\}_{j=1}^{m}$ are pairwise distinct
and simple. Given $\gamma>0$, the proper transfer functions
$P=n_P/d_P$ and $P_a=n_{P_a}/d_{P_a}$ satisfy
\begin{equation}
P(s)-P_a(s)
=
\gamma\,
\frac{d_P(-s)d_{P_a}(-s)}
     {d_P(s)d_{P_a}(s)}
\label{eq:allpass_error_identity}
\end{equation}
if and only if
\begin{subequations}\label{eq:numerator_interpolation}
\begin{align}
n_P(p_i)
&=
\gamma\,
\frac{d_P(-p_i)d_{P_a}(-p_i)}
     {d_{P_a}(p_i)},
&& i=1,\ldots,n,\\
n_{P_a}(q_j)
&=
-\gamma\,
\frac{d_P(-q_j)d_{P_a}(-q_j)}
     {d_P(q_j)},
&& j=1,\ldots,m,
\end{align}
\end{subequations}
and their high-frequency gains satisfy
\begin{equation}
P(\infty)-P_a(\infty)=(-1)^{n+m}\gamma.
\label{eq:high_frequency_gain_condition}
\end{equation}
For any specified value of either $P(\infty)$ or $P_a(\infty)$,
these conditions uniquely determine $n_P$ and $n_{P_a}$.
\end{proposition}

\begin{proof}
Clearing the denominators in \eqref{eq:allpass_error_identity} leads to
\begin{equation}
n_P(s)d_{P_a}(s)-n_{P_a}(s)d_P(s)
=\gamma d_P(-s)d_{P_a}(-s).
\label{eq:polynomial_error_identity}
\end{equation}
Evaluating \eqref{eq:polynomial_error_identity} at the roots $p_i$
and $q_j$ gives the interpolation conditions given in \eqref{eq:numerator_interpolation}. Equating the
coefficients of $s^{n+m}$ gives
\eqref{eq:high_frequency_gain_condition}.

Conversely, let 
\[
F(s)=n_P(s)d_{P_a}(s)-n_{P_a}(s)d_P(s)
      -\gamma d_P(-s)d_{P_a}(-s)~.
\]
The interpolation conditions
give $F(p_i)=F(q_j)=0$ for all $i$ and $j$. The gain condition cancels the coefficient of $s^{n+m}$ in $F$.
Hence $F$ is either identically zero or has degree at most $n+m-1$.
Since the interpolation conditions give $n+m$ distinct roots, only
$F\equiv0$ is possible. Thus, \eqref{eq:allpass_error_identity} holds.
\end{proof}

\FloatBarrier
\bibliographystyle{unsrtnat}
\bibliography{references}
\end{document}